\documentclass[11pt]{article}

\usepackage{hyperref,amsfonts,amsmath,amssymb,graphicx,bm,subfigure,a4wide,cite}

\numberwithin{equation}{section}

\begin{document}

\title{\textbf{Production of relic gravitational waves and the baryon asymmetry of the universe by random hypermagnetic fields}}

\author{Maxim Dvornikov\thanks{maxdvo@izmiran.ru}
\\
\small{\ Pushkov Institute of Terrestrial Magnetism, Ionosphere} \\
\small{and Radiowave Propagation (IZMIRAN),} \\
\small{108840 Moscow, Troitsk, Russia}}

\date{}

\maketitle

\begin{abstract}
We study the evolution of hypermagnetic fields (HMFs) in random plasma in the symmetric phase of the early universe. The system of kinetic equations for the spectra of the energy density and the helicity, as well as the particles asymmetries is derived. We also formulate the initial condition which involve the Kazantsev and Kolmogorov spectra of the seed HMFs. This system is solved numerically. We predict the energy spectrum of primeval gravitational waves which are produced by these HMFs. Additionally, the baryon asymmetry of the universe, generated by the lepton asymmetries, is obtained. These results allow us to constrain the strength of seed HMFs. 
\end{abstract}

\section{Introduction}

Cosmic magnetic fields, which permeate the universe, can exist even in the intergalactic space~\cite{NerVov10}. Such large scale magnetic fields cannot be produced by an astrophysical plasma motion. There are suggestions that cosmic magnetic fields proceed from the early universe~\cite{Vac21}. For example, usual Maxwell (electro-)magnetic field $A^\mu$ can be the remnant of the hypercharge field $Y^\mu = \cos\theta_\mathrm{W} A^\mu - \sin\theta_\mathrm{W} Z^\mu$, which existed before the electroweak phase transition (EWPT). Here, $\theta_\mathrm{W}$ is the Weinberg angle and $Z^\mu$ is the $Z$-boson field, which was massless before EWPT. During EWPT, which happens at $T_\mathrm{EW} = 100\,\text{GeV}$, $Y^\mu$ gives rise to present days photons, $A^\mu (T_\mathrm{EW}) = \cos\theta_\mathrm{W} Y^\mu (T_\mathrm{EW})$. The decomposition of the strength tensor of $Y^\mu$, $F_{\mu\nu} = (\mathbf{E}_\mathrm{Y},\mathbf{B}_\mathrm{Y})$, defines the hyperelectric $\mathbf{E}_\mathrm{Y}$ and hypermagnetic $\mathbf{B}_\mathrm{Y}$ fields (HMFs).

The dynamics of HMFs is affected by the asymmetries of fermions which possess nonzero hypercharges. It happens owing to the analogue of the chiral magnetic effect (the CME)~\cite{NieNin81}. In this situation, HMF is dynamo amplified and becomes unstable. The particle asymmetries also can be influenced by nonzero HMFs through the abelian anomaly (see, e.g., Ref.~\cite{Wei96}). These quantum phenomena can happen in the early universe before EWPT since all fermions are massless at that time.

The generation of lepton asymmetries by helical HMFs was noticed in Ref.~\cite{JoySha97} to result in the leptogenesis and, then, in the baryogenesis. This scenario was generalized in Ref.~\cite{KamLon16} to add the quarks asymmetries. The coexistence of HMFs and particle asymmetries was considered in Refs.~\cite{DvoSem12,DvoSem13,SemSmiSok16,AbbZad21,DvoSem21,Dvo23}. Other mechanisms for the production of the baryon asymmetry of the universe (BAU), including those based on the physics beyond the standard model, are reviewed in Ref.~\cite{BodBuc21}.

Evolving HMFs can generate both BAU and relic gravitational waves (GWs). Unlike BAU, which is produced by lepton asymmetries evolving together with HMFs, relic GWs stem directly from HMFs. Namely, the GW background appears when the energy-momentum tensor of HMFs is accounted for in the right hand side of the Einstein equation. Such a possibility was first considered in Ref.~\cite{KosMacKah02}. The production of primordial GWs by turbulent magnetic fields amplified by the CME was studied in Ref.~\cite{Bra21} basing on the numerical solution of the full set of the anomalous MHD equations. We studied the generation of relic GWs by random HMFs in Refs.~\cite{Dvo22,Dvo23}. Other possibilities to generate a GW background in the early universe, especially by modifying the General Relativity, are reviewed in Ref.~\cite{OdiOikMyr22}.

The studies of relic GWs are inspired by both the multiple direct detections of GWs produced by coalescing astrophysical objects~\cite{Abb21a} and a recent claim in Ref.~\cite{Arz20} that a stochastic GW background is observed. Moreover, several future GW telescopes, for instance, based in satellite~\cite{Eva21,Auc22} and underground~\cite{Mag20} facilities, are designed to probe stochastic relic GWs.

This work is organized as follows. We start with the study of the HMFs dynamics in Sec.~\ref{sec:EVOLHMF}, where we formulate the kinetic equations for the spectra and the asymmetries, as well as set up the initial condition. Then, in Sec.~\ref{sec:GW}, we consider the production of relic GWs basing on the obtained spectra of HMFs. The generation of BAU, which depends on the leptons asymmetries, is studied in Sec.~\ref{sec:BAU}. Finally, in Sec.~\ref{sec:CONCL}, we conclude.

\section{Dynamics of random HMFs\label{sec:EVOLHMF}}

We study the evolution of HMFs in the symmetric phase of the universe before EWPT when all particles are massless. In this situation, the divergence of an axial current of leptons is nonzero because of the abelian anomalies~\cite{Wei96},
\begin{equation}\label{adler}
  \partial_\mu j^\mu_5 =
  \partial_\mu 
  \langle
    \bar{\psi} \gamma^\mu \gamma^5 \psi
  \rangle =
  \frac{g^2}{2\pi^2}(\mathbf{E}_\mathrm{Y}\mathbf{B}_\mathrm{Y}),
\end{equation}
where $g$ is the hypercharge, $\gamma^\mu = (\gamma^0,\bm{\gamma})$ and $\gamma^5$ are the Dirac matrices. Moreover, a vector current of these leptons acquires the component along the HMF,
\begin{equation}\label{CME}
  \mathbf{j} = 
  \langle
    \bar{\psi} \bm{\gamma} \psi
  \rangle =
  \frac{g^2}{2\pi^2}\mu_5 \mathbf{B}_\mathrm{Y},
  \quad
  \mu_5 = \frac{1}{2}(\mu_\mathrm{R}-\mu_\mathrm{L}),
\end{equation}
where $\mu_\mathrm{R,L}$ are the chemical potentials of right and left fermions. In Eq.~\eqref{CME}, we show only the spatial components of the current. One can see in Eq.~\eqref{CME} that it is analogous to the CME~\cite{NieNin81}.

HMFs in the early universe are stochastic, i.e. $\langle \mathbf{B}_\mathrm{Y} \rangle = 0$. However, binary combinations of HMFs are nonzero. We deal mainly with the hypermagnetic energy density, $B_\mathrm{Y}^2/2$, and the hypermagnetic helicity, $\smallint (\mathbf{Y}\mathbf{B}_\mathrm{Y}) \mathrm{d}^3x$, where $\mathbf{Y}$ is the vector potential of the hypercharge field. The hypermagnetic helicity characterizes the topology of HMF~\cite{Par20}. In fact, it is more convenient to study the spectra of these quantities,
\begin{equation}\label{spec}
  \frac{\tilde{B}_{\mathrm{Y}}^{2}}{2}=\int\mathrm{d}\tilde{k}\tilde{\mathcal{E}}_{{\rm B_{\mathrm{Y}}}}(\tilde{k}),
  \quad
  \int\frac{\mathrm{d}^{3}x}{V}(\tilde{\mathbf{Y}}\tilde{\mathbf{B}}_{\mathrm{Y}})=
  \int\mathrm{d}\tilde{k}\tilde{\mathcal{H}}_{{\rm B_{\mathrm{Y}}}}(\tilde{k})
\end{equation}
where we use the conformal variables $\tilde{k}=k_{\mathrm{phys}}/T$ and $\tilde{\mathbf{B}}_{\mathrm{Y}} = \mathbf{B}_{\mathrm{Y}}/T^2$. Here $T$ is the plasma temperature.

The plasma motion also affects the evolution of HMFs. The most consistent way to account for the plasma influence is to solve the full set of (H)MHD equations, which includes the Navier-Stokes equation for the plasma velocity. This task can be implemented only numerically (see, e.g., Ref.~\cite{Bra21}). To build an analytical model for random HMFs, we use the concept of the (H)MHD turbulence, which involves the replacement of the plasma velocity with the Lorentz force~\cite{Sig02}. 

Taking into account these factors, we derive the system of equations for the spectra in Eq.~\eqref{spec} (see, e.g., Ref.~\cite{Dvo23}),
\begin{align}\label{eq:HMFsys}
  \frac{\partial\tilde{\mathcal{E}}_{{\rm B_{\mathrm{Y}}}}}{\partial\tilde{\eta}} & =
  -2\tilde{k}^{2}\eta_{\mathrm{eff}}\tilde{\mathcal{E}}_{{\rm B_{\mathrm{Y}}}}+
  \alpha_{\mathrm{eff}}\tilde{k}^{2}\tilde{\mathcal{H}}_{{\rm B_{\mathrm{Y}}}},
  \nonumber
  \\
  \frac{\partial\tilde{\mathcal{H}}_{{\rm B_{\mathrm{Y}}}}}{\partial\tilde{\eta}} & =
  -2\tilde{k}^{2}\eta_{\mathrm{eff}}\tilde{\mathcal{H}}_{{\rm B_{\mathrm{Y}}}}+
  4\alpha_{\mathrm{eff}}\tilde{\mathcal{E}}_{{\rm B_{\mathrm{Y}}}},
\end{align}
where $\tilde{\eta}=\tilde{M}_{\mathrm{Pl}}(T^{-1}-T_{\mathrm{RL}}^{-1})$ is the conformal time, $\tilde{M}_{\mathrm{Pl}}=M_{\mathrm{Pl}}/1.66\sqrt{g_{*}}$, $T_{\mathrm{RL}}=10\,\text{TeV}$ is the
temperature when left femions start to be produced,
$M_{\mathrm{Pl}}=1.2\times10^{19}\,\text{GeV}$ is the Planck mass,
$g_{*}=106.75$ is the number of the relativistic degrees of freedom before EWPT. The explicit form of the effective coefficients, $\eta_{\mathrm{eff}}$ and $\alpha_{\mathrm{eff}}$, which account for the CME in Eq.~\eqref{CME} and (H)MHD turbulence, can be found in Refs.~\cite{Cam07,Dvo23}.

Assuming that plasma in the early universe is uniform and integrating Eq.~\eqref{adler} over space, we get the system of the kinetic equations for the asymmetries of all types of leptons, in which we account for their decays and the sphaleron process. We also add the kinetic equation for the Higgs boson asymmetry. Finally, one has the system in the form,
\begin{align}\label{eq:asymsys}
  \frac{\mathrm{d}\xi_{e\mathrm{R}}}{\mathrm{d}\tilde{\eta}} & =
  -\frac{6\alpha'}{\pi}\int\mathrm{d}\tilde{k}
  \left(
    2\alpha_{\mathrm{eff}}\tilde{\mathcal{E}}_{{\rm B_{\mathrm{Y}}}}
    -\tilde{k}^{2}\eta_{\mathrm{eff}}\tilde{\mathcal{H}}_{{\rm B_{\mathrm{Y}}}}
  \right)-
  \Gamma(\xi_{e\mathrm{R}}-\xi_{e\mathrm{L}}+\xi_{0}),
  \nonumber
  \\
  \frac{\mathrm{d}\xi_{e\mathrm{L}}}{\mathrm{d}\tilde{\eta}} & =
  \frac{3\alpha'}{2\pi}\int\mathrm{d}\tilde{k}
  \left(
    2\alpha_{\mathrm{eff}}\tilde{\mathcal{E}}_{{\rm B_{\mathrm{Y}}}}
    -\tilde{k}^{2}\eta_{\mathrm{eff}}\tilde{\mathcal{H}}_{{\rm B_{\mathrm{Y}}}}
  \right)-
  \frac{\Gamma}{2}(\xi_{e\mathrm{L}}-\xi_{e\mathrm{R}}-\xi_{0})-\frac{\Gamma_{\mathrm{sph}}}{2}\xi_{e\mathrm{L}},
  \nonumber
  \\
  \frac{\mathrm{d}\xi_{0}}{\mathrm{d}\tilde{\eta}} & =
  -\frac{\Gamma}{2}(\xi_{e\mathrm{R}}+\xi_{0}-\xi_{e\mathrm{L}}),
\end{align}
where the asymmetries of right and left fermions, as well as of Higgs bosons are $\xi_{e\mathrm{R,L}}=6(n_{e\mathrm{R,L}}-n_{\bar{e}\mathrm{R,L}})/T^{3}$ and $\xi_{0}=3(n_{\varphi_{0}}-n_{\bar{\varphi}_{0}})/T^{3}$, respectively. The values of the rates $\Gamma$ and $\Gamma_\mathrm{sph}$, for the sphaleron process, are given in Ref.~\cite{DvoSem21}.

To integrate the system in Eqs.~\eqref{eq:HMFsys} and~\eqref{eq:asymsys} we should specify the initial condition. We suppose that the energy spectrum of a seed HMF at $T = T_{\mathrm{RL}}=10\,\text{TeV}$, when left leptons appear, consists of two parts. The infrared (IR) part is the Kazantsev spectrum $\propto \tilde{k}^{3/2}$~\cite{Kaz68}, whereas the ultraviolet part is the Kolmogorov spectrum $\propto \tilde{k}^{-5/3}$. The Kazantsev spectrum can correspond to reciprocal momenta greater than the horizon size~\cite{Bra17}. Two branches in the spectrum are glued at certain $\tilde{k}_\star < \tilde{k}_\mathrm{max}$, where $\tilde{k}_\mathrm{max}$ is the maximal momentum which is a free parameter of the system. Demanding the plasma electroneutrality, we get that $\tilde{k}_\mathrm{max}<0.1$~\cite{DvoSem21}. Finally, the energy spectrum of a seed HMF is
\begin{equation}\label{eq:seedR0}
  \tilde{\mathcal{E}}^{(0)}_{{\rm B_{\mathrm{Y}}}}(\tilde{k})=\frac{\tilde{B}_{0}^{2}}{3\tilde{k}_\mathrm{max}}
  \left[
    \frac{19}{15}-\left( \frac{\tilde{k}_\star}{\tilde{k}_\mathrm{max}} \right)^{2/3}
  \right]^{-1}
  \times
  \begin{cases}
    \tilde{k}_\mathrm{max} \tilde{k}_\star^{-5/2} \tilde{k}^{3/2}, & 0<\tilde{k}<\tilde{k}_\star,
    \\
    \tilde{k}_\mathrm{max} \tilde{k}_\star^{2/3} \tilde{k}^{-5/3}, & \tilde{k}_\star<\tilde{k}<\tilde{k}_\mathrm{max},
  \end{cases}
\end{equation}
where $\tilde{B}_{0}$ is the strength of a seed HMF.

The spectrum of a seed helicity density is $\mathcal{\tilde{H}}_{{\rm B_{\mathrm{Y}}}}^{(0)}(\tilde{k})=2q\mathcal{\tilde{E}}_{{\rm B_{\mathrm{Y}}}}^{(0)}(\tilde{k})/\tilde{k}$,
where $0\leq q\leq 1$ is the phenomenological parameter fixing the helicity
of a seed HMF. In our simulations, we take that $q=1$ assuming maximally helical HMFs.

Additionally to Eq.~\eqref{eq:seedR0}, we should set the initial asymmetries in
Eq.~(\ref{eq:asymsys}). We take that $\xi_{e\mathrm{L}}=\xi_{0}=0$
and $\xi_{e\mathrm{R}}=10^{-10}$ following Refs.~\cite{GioSha98,DvoSem13,DvoSem21,Dvo22,Dvo23}. It means that the main contribution to the dynamics of the system results from the  the right electrons component.

\begin{figure}
  \centering
  \subfigure[]
  {\label{fig:1a}
  \includegraphics[scale=.35]{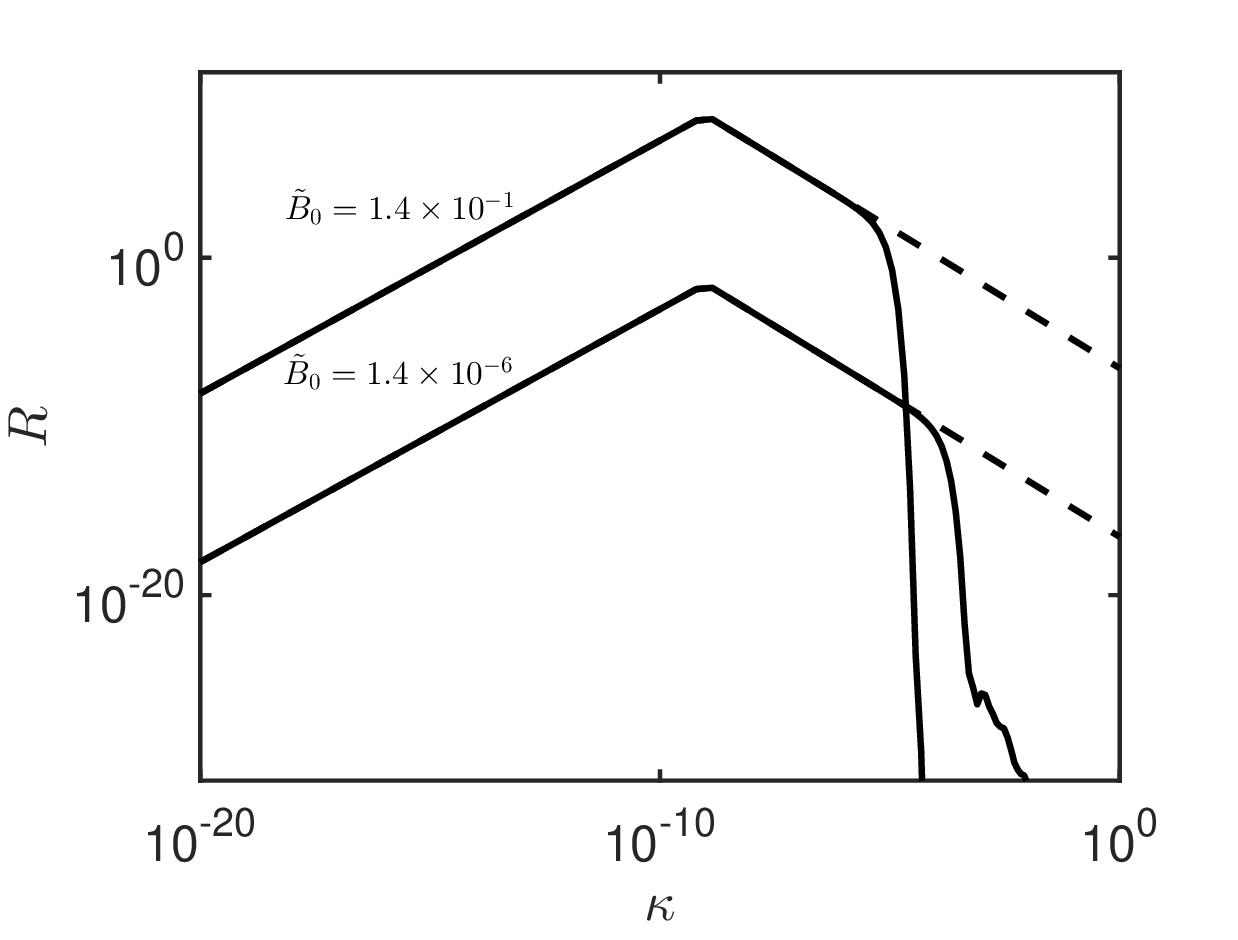}}
  \hskip-.5cm
  \subfigure[]
  {\label{fig:1b}
  \includegraphics[scale=.35]{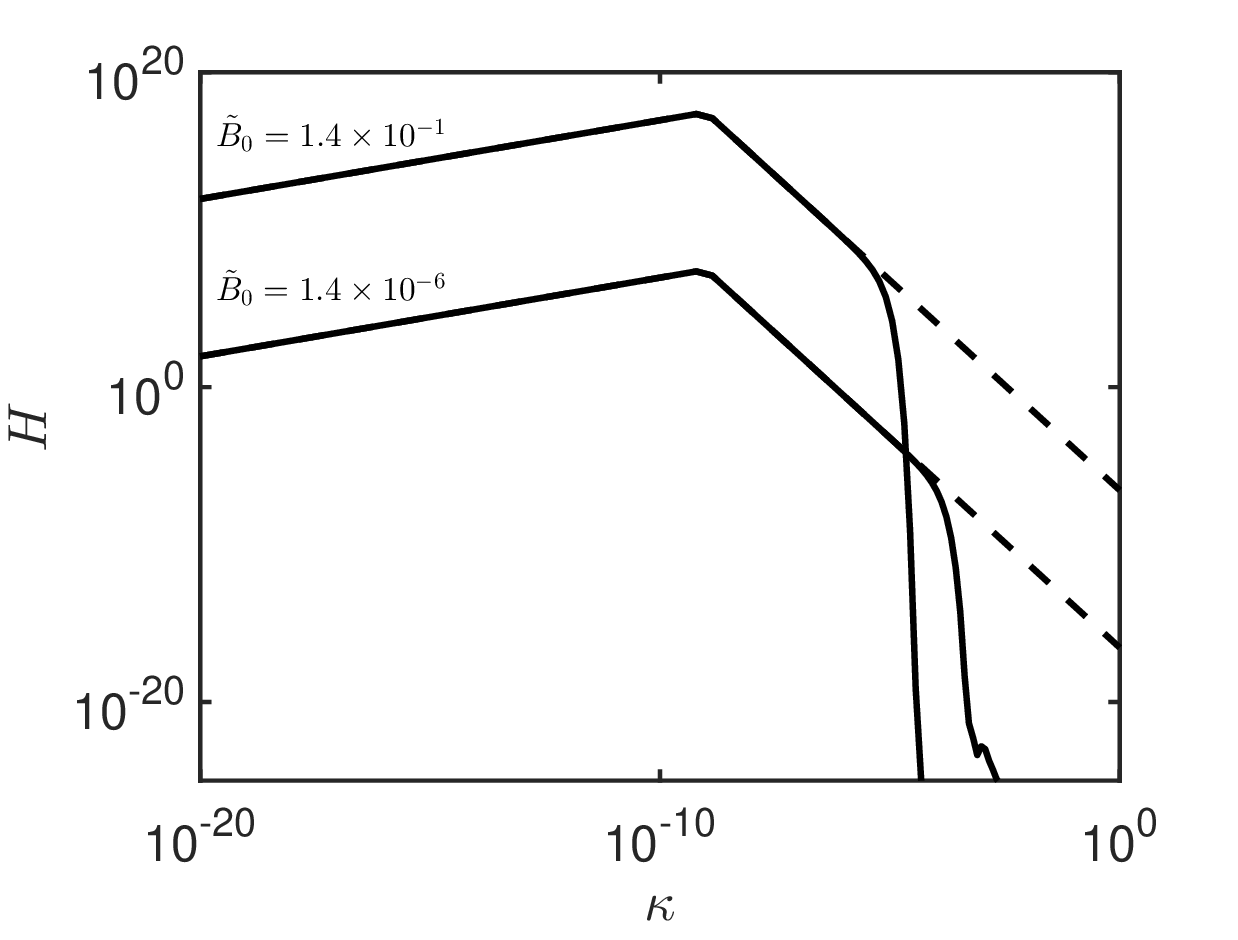}}
  \\
  \vskip-.3cm
  \subfigure[]
  {\label{fig:1c}
  \includegraphics[scale=.35]{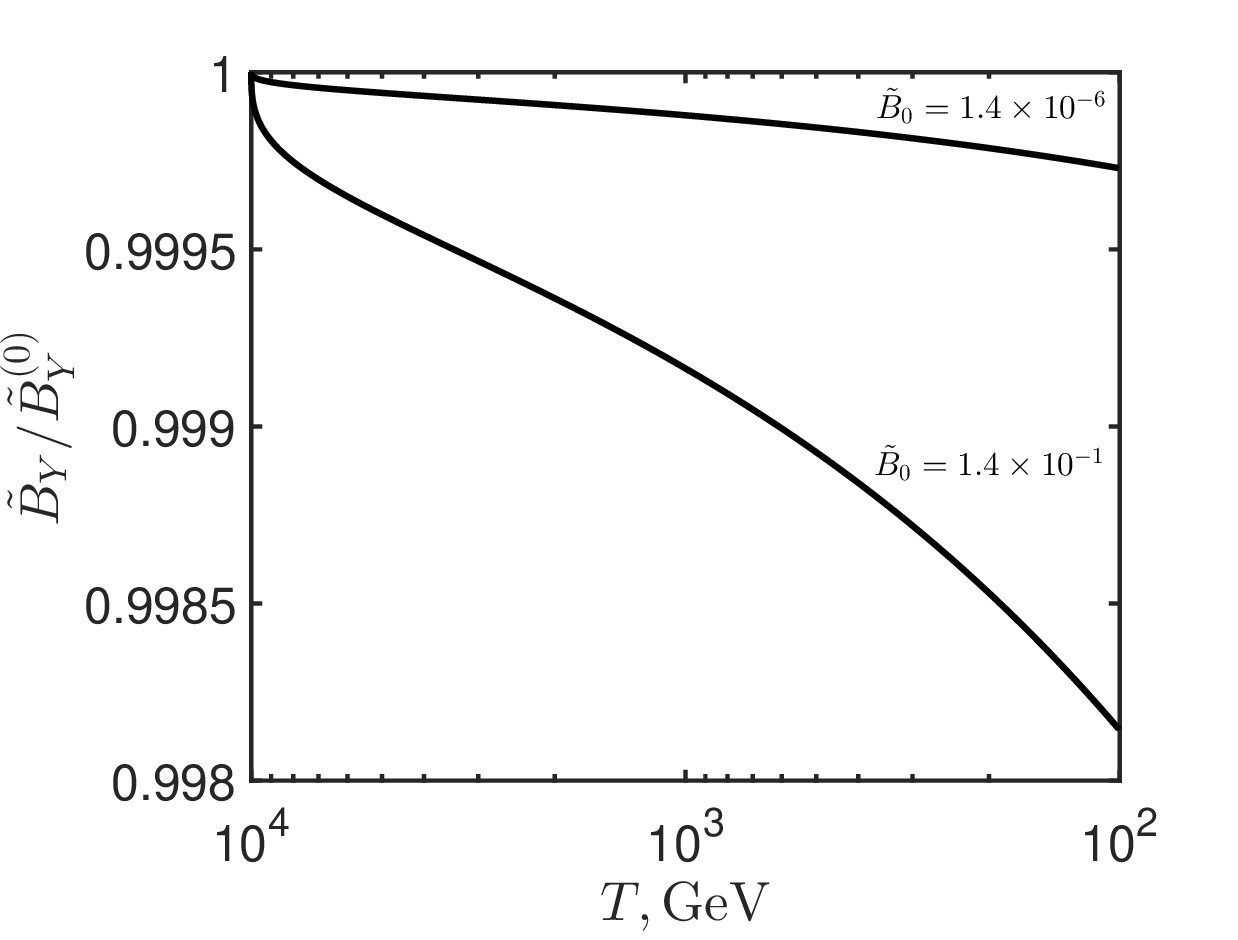}}
  \hskip-.5cm
  \subfigure[]
  {\label{fig:1d}
  \includegraphics[scale=.35]{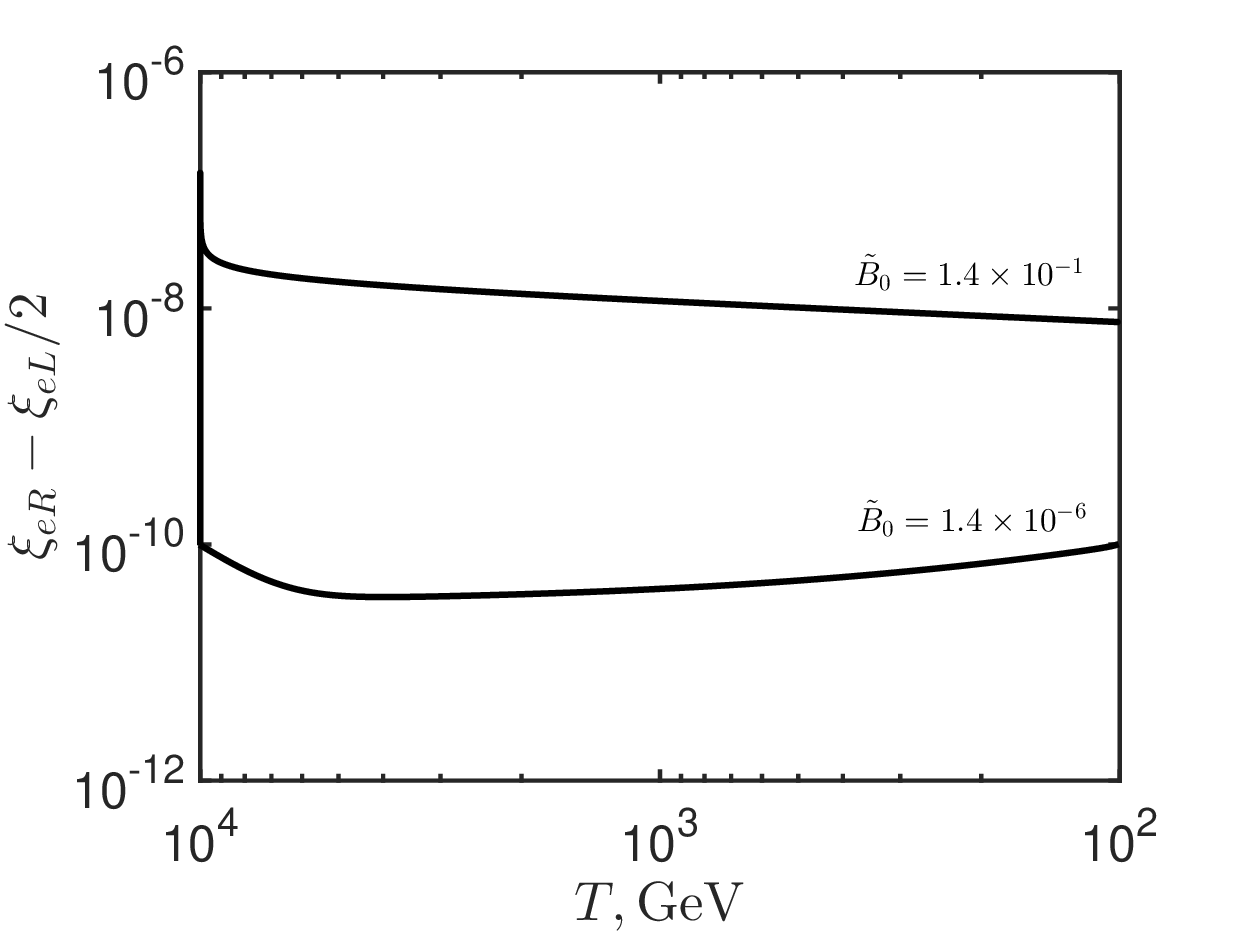}}
  \protect
\caption{The results of the numerical solution of Eqs.~\eqref{eq:HMFsys} and~\eqref{eq:asymsys}. We take that $\tilde{k}_{\mathrm{max}} = 10^{-3}$, $\tilde{k}_\star = 10^{-12}$, and $\xi_{e\mathrm{R}}^{(0)} = 10^{-10}$. All curves are plotted for $\tilde{B}_0 = 1.4\times 10^{-6}$ and $\tilde{B}_0 = 1.4\times 10^{-1}$. (a) The dimensionless spectrum of the energy density $R$ versus $\kappa$.  (b) The dimensionless spectrum of the helicity density $H$ versus $\kappa$. In panels (a) and (b), we also show the corresponding spectra on the seed HMFs by dashed lines. (c) The normalized HMF strength versus the plasma temperature. (d) The quantity $\xi_{e\mathrm{R}} -  \xi_{e\mathrm{L}}/2$ in the cooling universe.\label{fig:spectra}}
\end{figure}

Now, we show the results of the numerical solution of Eqs.~\eqref{eq:HMFsys} and~\eqref{eq:asymsys} for the chosen initial condition. First, in Figs.~\ref{fig:1a} and~\ref{fig:1b}, we demonstrate the evolution of the dimensionless spectra $R(\kappa) = 6\alpha'^{2}\tilde{\mathcal{E}}_{{\rm B_{\mathrm{Y}}}}(\tilde{k})/\pi^{2}\tilde{k}_{\mathrm{max}}$ and $H(\kappa) = 3\alpha'^{2}\tilde{\mathcal{H}}_{{\rm B_{\mathrm{Y}}}}(\tilde{k})/\pi^{2}$, where $\kappa = \tilde{k}/\tilde{k}_{\mathrm{max}}$ and $\alpha'$ is the analogue of the fine structure constant for the hypercharge field, from their initial values at $T=T_\mathrm{RL}$ (dashed lines) to EWPT (solid lines). The spectra are shown for two seed strengths, $\tilde{B}_0 = 1.4\times 10^{-6}$ and $\tilde{B}_0 = 1.4\times 10^{-1}$.

We can see in Figs.~\ref{fig:1a} and~\ref{fig:1b} that the spectra are the subject of mainly the magnetic diffusion at great $\kappa$. It is owing to the great $\eta_\mathrm{eff}$ in Eq.~\eqref{eq:HMFsys}, which is enhanced by the (H)MHD turbulence contribution. The irregular parts of the spectra at the very great $\kappa \lesssim 1$ are because of the finite accuracy of numerical simulations. To see more clearly that HMF decays, in Fig.~\ref{fig:1c}, we show its time evolution in the early universe cooling from $T_\mathrm{RL}$ down to EWPT. The curves are plotted again for two different strengths of a seed HMF.

Finally, in Fig.~\ref{fig:1d}, we represent the evolution of the chiral $\alpha$-parameter, $\alpha_\mathrm{Y} \propto \xi_{e\mathrm{R}} - \xi_{e\mathrm{L}}/2$ (see, Refs.~\cite{DvoSem21,Dvo22,Dvo23}). It is important for the BAU generation, which is discussed in Sec.~\ref{sec:BAU}. It should be noted that, qualitatively, the behavior of HMFs and the asymmetries, in case when we take the Kazantsev spectrum of the seed field in Eq.~\eqref{eq:seedR0}, is similar to that for the Batchelor IR part of the spectrum studied in Ref.~\cite{Dvo23}.

\section{Generation of relic GWs by stochastic HMFs\label{sec:GW}}

A gravitational field, corresponding to the metric $g_{\mu\nu}$, evolves under the influence of external HMFs as (see, e.g., Ref.~\cite{Wei20})
\begin{equation}\label{Einstein}
  R_{\mu\nu} - \frac{1}{2} g_{\mu\nu}R = 8\pi G T_{\mu\nu},
\end{equation}
where $R_{\mu\nu}$ is the Ricci tensor, $R$ is the scalar curvature, $G$ is the Newton constant, and the spatial components of the energy-momentum tensor $T_{\mu\nu}$ are
\begin{equation}\label{eq:Tij}
  T_{ij}=-\frac{1}{a^{2}}\left(B_{\mathrm{Y}i}^{(c)}B_{\mathrm{Y}j}^{(c)}-\frac{1}{2}\delta_{ij}B_{\mathrm{Y}}^{(c)2}\right).
\end{equation}
Here $\mathbf{B}_{\mathrm{Y}}^{(c)}=a^{2}\mathbf{B}_{\mathrm{Y}}$
is the conformal HMF and $a$ is the scale factor in the Friedmann--Robertson--Walker metric, $\bar{g}_{\mu\nu}=\text{diag}(1,-a^{2},-a^{2},-a^{2})$. 

Considering the tensor perturbations of the metric in Eq.~\eqref{Einstein}, $g_{\mu\nu} = \bar{g}_{\mu\nu} + h_{\mu\nu}$, and using the transverse-traceless gauge~\cite{Wei20}, we get the spectrum of the energy density of stochastic GWs~\cite{Dvo22}
\begin{align}\label{eq:rhoGWisotr}
  \rho_{\mathrm{GW}}^{(c)}(k,\eta)= & \frac{t_{\text{Univ}}^{2}G}{4k^{3}\pi^{2}}\eta
  \int_{0}^{\eta}\frac{\mathrm{d}\xi}{(\eta_{0}+\xi)^{2}}
  \int_{0}^{\infty}\frac{\mathrm{d}q}{q^{3}}\int_{|k-q|}^{k+q}\frac{\mathrm{d}p}{p^{3}}
  \nonumber
  \\
  & \times
  \big\{
    [4k^{2}q^{2}+(k^{2}+q^{2}-p^{2})^{2}][4k^{2}p^{2}+(k^{2}-q^{2}+p^{2})^{2}]
    \rho_{\mathrm{Y}}^{(c)}(q,\xi)\rho_{\mathrm{Y}}^{(c)}(p,\xi)
    \nonumber
    \\
    & +
    4k^{2}q^{2}p^{2}(k^{2}+q^{2}-p^{2})(k^{2}-q^{2}+p^{2})
    h_{\mathrm{Y}}^{(c)}(q,\xi)h_{\mathrm{Y}}^{(c)}(p,\xi)
  \big\}
\end{align}
Here $\rho_{\mathrm{Y}}^{(c)}(k,\eta)$ and $h_{\mathrm{Y}}^{(c)}(k,\eta)$
are the conformal spectra of HMFs
energy and helicity, which are related to the spectra, given in
Sec.~\ref{sec:EVOLHMF}, by $\rho_{\mathrm{Y}}^{(c)}(k,\xi)=\tilde{\mathcal{E}}_{{\rm B_{\mathrm{Y}}}}(\tilde{k},\tilde{\eta})T_{0}^{3}$
and $h_{\mathrm{Y}}^{(c)}(k,\xi)=\tilde{\mathcal{H}}_{{\rm B_{\mathrm{Y}}}}(\tilde{k},\tilde{\eta})T_{0}^{2}$,
where $T_{0}=2.7\,\text{K}$ is the current temperature of the cosmic
microwave background radiation. The dimensional conformal time $\eta$
and the conformal momentum $k$ in Eq.~(\ref{eq:rhoGWisotr}) are $\eta=(2t_{\mathrm{Univ}}T_{0}/\tilde{M}_{\mathrm{Pl}})\tilde{\eta}$
and $k=T_{0}\tilde{k}$, where $t_{\mathrm{Univ}}=1.4\times10^{10}\,\text{yr}$
is the age of the universe. In Eq.~(\ref{eq:rhoGWisotr}),
the parameter $\eta_{0}=2t_{\mathrm{Univ}}T_{0}/T_{\mathrm{RL}}$.

\begin{figure}
  \centering
  \includegraphics[scale=.35]{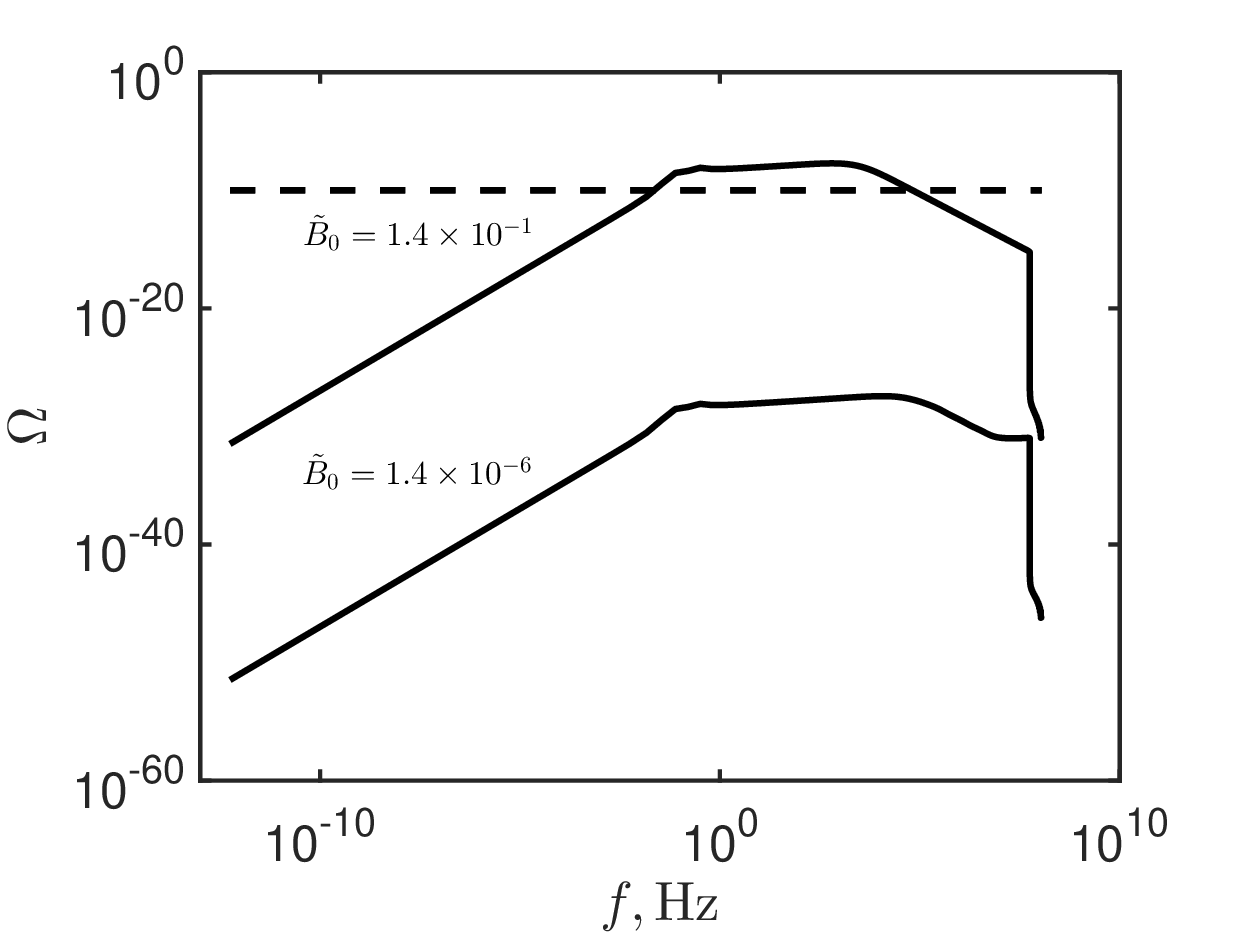}
  \protect
\caption{The observed energy spectrum of relic GWs based on the behavior of random HMFs in Sec.~\ref{sec:EVOLHMF}. The parameters of the system are the same as in Fig.~\ref{fig:spectra}. By the dashed line, we show the observational upper bound on $\Omega$~\cite{Abb21}.\label{fig:gw}}
\end{figure}

Instead of $\rho_{\mathrm{GW}}^{(c)}(k,\eta)$ in Eq.~\eqref{eq:rhoGWisotr}, a GW detector measures the quantity,
\begin{equation}\label{eq:Omegadef}
  \Omega(f)=\frac{1}{\rho_\mathrm{crit}}\frac{\text{d}\rho_\mathrm{GW}}{\text{d}\ln f}=
  \frac{f\rho_{\mathrm{GW}}(f)}{\rho_{\mathrm{crit}}},
\end{equation}
where $\rho_{\mathrm{crit}}=0.53\times10^{-5}\,\text{GeV}\cdot\text{cm}^{-3}$
is the critical energy density of the universe, and $f$ is the frequency measured in Hz. In Fig.~\ref{fig:gw}, we show $\Omega(f)$ in Eq.~\eqref{eq:Omegadef} basing on the spectra of HMFs found in Sec.~\ref{sec:EVOLHMF}. As in Fig.~\ref{fig:spectra}, we plot $\Omega$ for two different seed strengths, $\tilde{B}_0 = 1.4\times 10^{-6}$ and $\tilde{B}_0 = 1.4\times 10^{-1}$.

Figure~\ref{fig:gw} corresponds to $T=T_\mathrm{EW}=100\,\text{GeV}$. However, since we deal with the conformal energy density of GWs, the same GW signal is observed nowadays. In Fig.~\ref{fig:gw}, we also plot by the dashed line the constraint on the stochastic GW background by the LIGO-Virgo-KAGRA collaborations in Ref.~\cite{Abb21}. The value $\Omega_\mathrm{obs} \sim 10^{-10}$ is established in the Hz--kHz frequency band. Thus, using Fig.~\ref{fig:gw}, we put the upper bound on the strength of a seed HMF, $\tilde{B}_0 \lesssim 0.14$.

\section{Production of BAU by evolving asymmetries\label{sec:BAU}}

We studied both the spectra of HMFs and particle asymmetries in Sec.~\ref{sec:EVOLHMF}. The spectra are applied in Sec.~\ref{sec:GW} to predict the GW background generation. Now, we use the lepton asymmetries to describe the production of BAU.

Basing on the 't~Hooft conservation law, $B-L=\text{const}$, we get that the value of BAU depends on the asymmetries of right and left leptons $\xi_{e\mathrm{R,L}}$~\cite{DvoSem21},
\begin{align}\label{eq:BAUgen}
  \text{BAU}(\tilde{\eta})= & \frac{n_{\mathrm{B}}-n_{\bar{\mathrm{B}}}}{s}=
  5.3\times10^{-3}\int_{0}^{\tilde{\eta}}\mathrm{d}\tilde{\eta}'
  \notag
  \\
  & \times
  \left\{
    \frac{{\rm d}\xi_{e\mathrm{R}}(\tilde{\eta}')}{{\rm d}\tilde{\eta}'}+
    \Gamma(\tilde{\eta}')[\xi_{e\mathrm{R}}(\tilde{\eta}')-\xi_{e\mathrm{L}}(\tilde{\eta}')]
  \right\} -
  \frac{6\times10^{7}}{\tilde{\eta}_{\mathrm{EW}}}
  \int_{0}^{\tilde{\eta}}\xi_{e\mathrm{L}}(\tilde{\eta}')\mathrm{d}\tilde{\eta}',
\end{align}
where $n_{\mathrm{B},\bar{\mathrm{B}}}$ are number densities of baryons
and antibaryons, $s$ is the entropy density, and $\tilde{\eta}_{\mathrm{EW}}$ is the dimensionless conformal time corresponding to EWPT. We assume in Eq.~\eqref{eq:BAUgen} that $\text{BAU}=0$ at $T=T_{\mathrm{RL}}$.

\begin{figure}
  \centering
  \includegraphics[scale=.35]{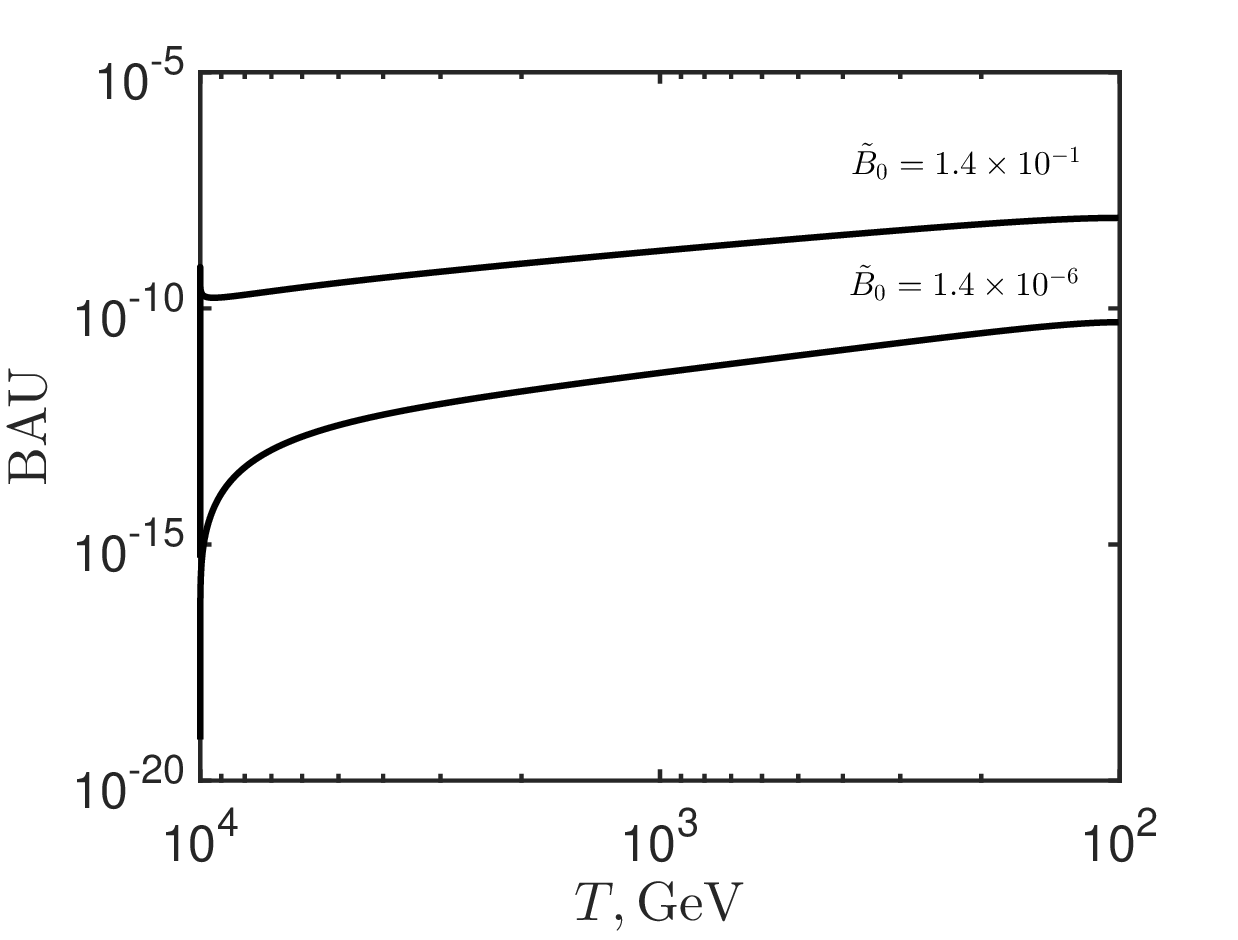}
  \protect
\caption{BAU in the early universe cooling from $T_\mathrm{RL}$ down to EWPT. We use the same parameters as in Figs.~\ref{fig:spectra} and~\ref{fig:gw}.\label{fig:bau}}
\end{figure}

BAU is shown in Fig.~\ref{fig:bau} for different strengths of a seed HMF. We can see in Fig.~\ref{fig:bau}, that the curve corresponding to $\tilde{B}_0 = 1.4\times 10^{-1}$ results in an excessive BAU. The weaker seed HMF with $\tilde{B}_0 = 1.4\times 10^{-6}$ leads to BAU comparable with the observed value $\text{BAU}_\mathrm{obs}\sim 10^{-10}$. Thus, we establish a stronger constraint on the strength of a seed HMF, $\tilde{B}_0 \lesssim 1.4\times 10^{-6}$. This result is analogous to that obtained in Ref.~\cite{Dvo23}.

If we compare BAU at EWPT in Fig.~\ref{fig:bau} with the value of $\xi_{e\mathrm{R}} -  \xi_{e\mathrm{L}}/2$ at $T=100\,\text{GeV}$ in Fig.~\ref{fig:1d}, we can see that BAU is determined mainly by the evolution of right electrons. This fact was mentioned first in Ref.~\cite{GioSha98}. We used this feature in the series of our works~\cite{DvoSem13,DvoSem21,Dvo23}.

\section{Conclusion\label{sec:CONCL}}

The present work has been devoted to the study of the evolution of HMFs in the symmetric phase of the early universe before EWPT. The dynamics of HMFs accounts for the analogues of the MHD turbulence and the CME in the presence of nonzero asymmetries of leptons. To close the system we describe the behavior of the asymmetries in HMFs taking into account the abelian anomalies. We deal with the binary combinations of HMFs, like the energy density and the helicity, since we consider random fields in the primordial plasma. The evolution of HMFs is tracked from $T_\mathrm{RL} = 10\,\text{TeV}$, when left fermions start to be produced, to EWPT which happens at $T_\mathrm{EW} = 100\,\text{GeV}$.

In Sec.~\ref{sec:EVOLHMF}, we have formulated the dynamics of HMFs and specified the initial condition. We have supposed that the spectrum of seed HMFs has the Kazantsev part $\propto \tilde{k}^{3/2}$ for small $\tilde{k}$ and the Kolmogorov part $\propto \tilde{k}^{-5/3}$ for great ones. The consideration of the Kazantsev spectrum is the advance compared to Ref.~\cite{Dvo23} where the Batchelor spectrum was used. Recently, the possibility to distinguish between the Kazantsev and Batchelor spectra in galactic magnetic fields was studied in Ref.~\cite{Bra22}.

Then, in Sec.~\ref{sec:GW}, we have studied the production of relic GWs by random HMFs. We have derived the spectrum of the conformal energy density of GWs. On the basis of this result, we have computed the function $\Omega$, which is measured by GW telescopes, at EWPT. We have compared this finding with the observational constraint in Ref.~\cite{Abb21} and established the upper bound on the strength of the seed HMF.

Finally, in Sec.~\ref{sec:BAU}, we have studied the generation of BAU on the basis of asymmetries of right and left leptons. BAU has been found to reach the observed value $\sim 10^{-10}$ if seed HMFs are constrained by a smaller value compared to the case of primeval GWs studied in Sec.~\ref{sec:GW}. It should be noted that the constraint on the strength of HMFs obtained in Sec.~\ref{sec:BAU} is consistent with the upper bounds established in Ref.~\cite{CheSchTru94}, based on the Big Bang nucleosynthesis consideration, and in Ref.~\cite{Kam20}, where the baryon isocurvature was discussed.

We discuss random fields in the early universe. They are considered as stochastic
processes. Strictly speaking, the mean value of a physical quantity should be computed basing on
a statistical ensemble. However, this procedure cannot be implemented in practice since one has
to consider multiple copies of the universe. However, using the ergodic hypothesis, we can
compute the mean value as a time averaging of a stochastic process. Such a quantity coincides
with the mean value calculated using the statistical averaging.

\end{document}